\documentstyle[preprint,aps,epsfig]{revtex} 
\newcommand{\be}[1]{\begin{equation} \label{(#1)}} 
\newcommand{\ee}{\end{equation}}  
\newcommand{\ba}[1]{\begin{eqnarray} \label{(#1)}} 
\newcommand{\ea}{\end{eqnarray}} 
\newcommand{\nn}{\nonumber}

\def\m{$\mu^--e^-$}

\tightenlines
 
\begin{document}

\title{Scalar meson mediated nuclear $\mu^--e^-$ conversion}

\author{Amand \ Faessler \footnotemark[1], 
Thomas \ Gutsche \footnotemark[1], 
Sergey \ Kovalenko \footnotemark[2], \\
Valery \ E. \ Lyubovitskij \footnotemark[1], 
Ivan \ Schmidt \footnotemark[2] 
\vspace*{0.4\baselineskip}}
\address{
\footnotemark[1] 
Institut f\"ur Theoretische Physik, Universit\"at T\"ubingen, \\
Auf der Morgenstelle 14, D-72076 T\"ubingen, Germany 
\vspace*{0.2\baselineskip}\\
\footnotemark[2]
Departamento de F\'\i sica, Universidad
T\'ecnica Federico Santa Mar\'\i a, \\ 
Casilla 110-V, Valpara\'\i so, Chile
\vspace*{0.3\baselineskip}\\}

\date{\today}
 
\maketitle 
 
\begin{abstract}
We study the nuclear $\mu^--e^-$ conversion in the general 
framework of the effective Lagrangian approach without referring to any 
specific realization of the physics beyond the standard model (SM) 
responsible for lepton flavor violation (LFV). 
We analyze the role of scalar meson exchange between the lepton and 
nucleon currents and show its relevance for the coherent channel of 
\m conversion. We show that this mechanism introduces 
modifications in the predicted \m conversion rates in comparison with 
the conventional direct nucleon mechanism, based on the contact type 
interactions of the nucleon currents with the LFV leptonic current.
We derive from the experimental data lower limits on the mass scales of 
the generic LFV lepton-quark contact terms and demonstrate that they are 
more stringent than the similar limits existing in the literature.

\vskip .3cm
 

\end{abstract}

\newpage 

\section{Introduction}

The study of lepton flavor violating (LFV) processes offers
a good opportunity for shedding light on the possible physics 
beyond the Standard Model (SM). 
Muon-to-electron ($\mu^--e^-$) conversion in nuclei 
\begin{eqnarray} 
\mu^- + (A,Z) \longrightarrow  e^- \,+\,(A,Z)^\ast 
\label{I.1} 
\end{eqnarray} 
is commonly recognized as one of the most sensitive probes of lepton 
flavor violation and of the related physics behind it (for reviews, 
see~\cite{Kosmas:1993ch,mu-e theory-exp}).

At present, on the experimental side there is one running \m conversion 
experiment, SINDRUM II~\cite{Honecker:1996zf}, and two planned ones, 
MECO~\cite{Molzon,MECO} and PRIME~\cite{PRIME}. 
So far this LFV process has not been observed and experimental results 
correspond to the upper limits on the \m conversion branching ratio
\begin{eqnarray}\label{Ti} 
&&R_{\mu e}^{A} = \frac{\Gamma(\mu^- + (A,Z) \rightarrow e^- + 
(A,Z))} {\Gamma(\mu^- + (A,Z)\rightarrow \nu_{\mu} + (A,Z-1))}\,.  
\end{eqnarray} 
The current and expected limits from the above mentioned experiments 
are presented in Table~I.
As is known from previous studies (see, for instance, 
Ref.~\cite{mu-e theory-exp,Faessler:2004ea} and references therein)
and will be discussed later in the present paper, these experimental 
bounds allow to set stringent limits on 
the mechanisms of \m conversion and the underlying theories of LFV. 

The theoretical studies of \m conversion, presented in the literature, 
cover various aspects of this LFV process, elaborating adequate treatment 
of the structure effects~\cite{Kosmas:1993ch,Kosmas:2001mv,Faessler:pn} of 
the nucleus participating in the reaction and, considering underlying 
mechanisms of LFV at the level of quarks within different scenarios of 
physics beyond the SM (see~\cite{mu-e theory-exp} and references therein). 

In general the \m conversion mechanisms can be classified as photonic and 
non-photonic. In the former case the \m conversion is mediated by the photon 
exchange between the LFV leptonic vertex and the ordinary electromagnetic 
nuclear vertex. The non-photonic mechanisms are based on the 4-fermion 
contact lepton-quark LFV interactions. These two categories of mechanisms 
differ significantly from each other since they receive different 
contributions from the new physics and require different treatment of the 
effects of the nucleon and the nuclear structure. 

In the present paper we continue studying the non-photonic meson exchange 
mechanism of \m conversion. Previously~\cite{Faessler:2004ea,Faessler:2004jt} 
we analyzed the vector-meson mediation of the \m conversion. 
The contribution of this mechanism to the coherent \m \, conversion 
results in some important issues for the physics beyond the SM absent in 
the case of the conventional direct lepton-nucleon interaction. Here, 
we extend our analysis to the scalar-meson exchange mechanism, which 
completes the study of meson exchange contributions to the coherent mode 
of the \m conversion \footnote{Pseudoscalar and axialvector mesons do not 
contribute to the coherent \m conversion.}.

\section{General Framework}

The effective Lagrangian ${\cal L}_{eff}^{lq}$ describing 
the coherent \m conversion at the quark level can be written in the 
form~\cite{Kosmas:2001mv,Faessler:2004jt}
\begin{eqnarray} 
{\cal L}_{eff}^{lq}\ =\  \frac{1}{\Lambda_{LFV}^2}  
\left[(\eta_{VV}^{q} j_{\mu}^V\ + \eta_{AV}^{q} 
j_{\mu}^A )J_{q}^{V\mu} + 
(\eta_{SS}^{q} j^S\ + \eta_{PS}^{q} j^P\ )J_{q}^{S}\right], 
\label{eff-q}
\end{eqnarray} 
where the lepton and quark currents are defined as:  
\begin{eqnarray}\label{lepton-currents}
&&j_{\mu}^V = \bar e \gamma_{\mu} \mu\,, \,\,\,  
j_{\mu}^A = \bar e \gamma_{\mu} \gamma_{5} \mu\,, \,\,\,  
j^S = \bar e \ \mu\,, \\ \nonumber
&& j^P = \bar e \gamma_{5} \mu\,, \,\,\,  
J_{q}^{V\mu} = \bar q \gamma^{\mu} q\,, \,\,\, 
J_{q}^{S} = \bar q \ q \,.
\end{eqnarray}   
In Eqs.~(\ref{eff-q}) and (\ref{lepton-currents}) the summation runs 
over all the quark species $q= \{u,d,s,b,c,t\}$.
The parameter $\Lambda_{LFV}$ with the dimension of mass is the 
characteristic high energy scale of lepton flavor violation attributed 
to new physics. The dimensionless LFV parameters $\eta^{q}$ in 
Eq.~(\ref{eff-q}) depend on a concrete LFV model. In the present study we 
treat these parameters as phenomenological parameters to be constrained from 
the experiment.

The quark level Lagrangian (\ref{eff-q}) generates the effective 
lepton-nucleon interactions that can be specified in terms of 
an effective Lagrangian on the nucleon level 
\begin{eqnarray} 
\hspace*{-.5cm} 
{\cal L}_{eff}^{lN} =   \frac{1}{\Lambda_{LFV}^2} 
\left[j_{\mu}^a (\alpha_{aV}^{(0)} J^{V\mu \, (0)} + 
\alpha_{aV}^{(3)} J^{V\mu \, (3)}) + j^b (\alpha_{bS}^{(0)} 
J^{S \, (0)} + \alpha_{bS}^{(3)} J^{S \, (3)})\right]\,. 
\label{eff-N} 
\end{eqnarray}  
Here, the isoscalar $J^{(0)}$ and isovector $J^{(3)}$ 
nucleon currents are 
$J^{V\mu \, (k)} = \bar N \, \gamma^\mu \, \tau^k \,  N\,$  and  
$J^{S \, (k)} = \bar N \, \tau^k \, N\,,$
where $N$ is the nucleon isospin doublet, $ k = 0,3 $ and 
$\tau_0\equiv\hat I$. 
The summation over the double indices 
$a = V,A$ and $b = S,P$ is implied in~(\ref{eff-N}).   
The nucleon Lagrangian~(\ref{eff-N}) is the basis 
for the derivation of the nuclear transition operators. 

Naturally, the Lagrangian in terms of effective nucleon 
fields~(\ref{eff-N}) is equivalent to the quark level 
Lagrangian~(\ref{eff-q}). The former Lagrangian is supposed to appear 
after the hadronization from the quark level Lagrangian~(\ref{eff-q}) 
and, therefore, must correspond to the same order $1/\Lambda_{LFV}^{2}$ 
in inverse powers of the LFV scale. 

To make a bridge between the underlying LFV physics and \m observables 
one needs to relate the lepton-nucleon LFV parameters $\alpha$ in 
Eq.~(\ref{eff-N}) to the lepton-quark LFV parameters $\eta$ in 
Eq.~(\ref{eff-q}). This implies a certain hadronization prescription 
which specifies the way in which the effect of quarks is simulated by 
hadrons. In the absence of a true theory of hadronization we rely on 
some reasonable assumptions and models. 
In Refs.~\cite{Faessler:2004ea,Faessler:2004jt} 
we considered the two mechanisms of nuclear $\mu^- - e^-$ conversion: 
direct nucleon mechanism and vector-meson exchange between nucleon and 
lepton currents. We found that the vector-meson exchange plays an 
important role in the coherent muon-electron conversion. Now, we extend 
this analysis to the scalar sector of the LFV Lagrangians in 
Eqs.~(\ref{eff-q}) and (\ref{eff-N}). 

As in the case of the vector currents, considered 
in Refs.~\cite{Faessler:2004ea,Faessler:2004jt}, here 
we distinguish the following two hadronization mechanism. 
The first one is the direct embedding of the quark currents into the 
nucleon (Fig.1a), which we call direct nucleon mechanism (DNM).  
The second mechanism consists of two stages (Fig.1b). First, the   
quark currents are embedded into the interpolating scalar meson fields which
then interact with the nucleon currents. We call this possibility 
meson-exchange mechanism (MEM). 

In general one expects all the mechanisms to contribute 
to the coupling constants $\alpha$ in Eq.~(\ref{eff-N}). However, 
at present the relative amplitudes of different mechanisms are unknown. 
In view of this problem we assume for the first approximation, that only one  
mechanism is operative and estimate its contribution to the process 
in question. This allows us to evaluate the importance of a specific mechanism.

Let us update the contribution of the direct nucleon mechanism derived 
in Ref.~\cite{Kosmas:2001mv}. The relation between the quark-lepton and 
nucleon-lepton LFV parameters in Eqs.~(\ref{eff-q}) and~(\ref{eff-N}) 
takes in this case the form 
\begin{eqnarray} 
\label{alpha}
\hspace*{-.65cm}
&&\alpha_{bS[DNM]}^{(3)} = \frac{1}{2}\eta_{bS}^{(3)}
(G_{S}^{u} - G_{S}^{d}),
\\ \nonumber 
&&\alpha_{bS[DNM]}^{(0)}  = \frac{1}{2}\eta_{bS}^{(0)}
(G_{S}^{u} + G_{S}^{d}) + \eta_{bS}^{s} G_{S}^{s},
\end{eqnarray} 
where $b=S,P$ and $\eta^{(0,3)}=\eta^{u}\pm\eta^{d}$ are the isoscalar and 
isovector quark couplings. The form factors $G_{S}^{q}$ are related to 
the scalar condensates in the nucleon~\cite{Inoue:2003bk}  
\begin{eqnarray}\label{mat-el1} 
&&\langle p|\bar{u}\ u|p\rangle = 
G_{S}^{u} \bar{p}\ p,  \ \ \  
\langle p|\bar{d}\ d|p\rangle = 
G_{S}^{d} \bar{p}\ p,  \ \ \ 
\langle p|\bar{s}\ s|p\rangle = 
G_{S}^{s} \bar{p}\ p,  
\\ \nn 
&&\langle n|\bar{u}\ u|n\rangle =  
G_{S}^{d} \bar{n}\ n, \ \ \  
\langle n|\bar{d}\ d|n\rangle = 
G_{S}^{u} \bar{n}\ n, \ \ \ 
\langle n|\bar{s}\ s|n\rangle = 
G_{S}^{s} \bar{n}\ n .  \nn 
\end{eqnarray}
Since the maximal momentum transfer $q$ in $\mu^- -e^-$ conversion is 
much smaller than the typical scale of the nucleon structure 
we can safely neglect the $q^2$-dependence of the nucleon form factors 
$G_{S}^{q}$. At $q^2=0$ these form factors are related to the corresponding 
meson-nucleon sigma-terms: 
\begin{eqnarray}\label{sigma}
\sigma_{\pi N} = \hat m  \, [ G_{S}^{u} + G_{S}^{d} ]\,, \,\,\, 
\sigma_{K N}^{I=1} &=& \frac{\hat m  + m_s}{4} \, [ G_{S}^{u} - G_{S}^{d} ]
\,, \,\,\, 
y_N = \frac{2 G_{S}^{s}}{G_{S}^{u} + G_{S}^{d}},
\end{eqnarray}
where $\hat m=(m_u+m_d)/2$ and $m_s$ are the masses of current quarks; 
$y_N$ is the strangeness of the nucleon. For these parameters we use the 
following values~\cite{Gasser:1982ap,Gasser:1990ce} in our analysis 
\begin{equation}
\hat m = 7 \;{\rm MeV},\; m_s/\hat m=25, \; y_N = 0.2\, .
\end{equation}
The canonical value of the $\pi N$ sigma term 
$\sigma_{\pi N} = 45 \pm 8$ MeV~\cite{Gasser:1990ce} was originally 
extracted from the dispersional analysis of $\pi N$ scattering data
taking into account chiral symmetry constraints. In particular,
the value of the sigma-term, $\sigma_{\pi N} = 45 \pm 8$ MeV, has been 
deduced from the analysis of two quantities: 
$\sigma_{\pi N}(t=2M_{\pi}^2) = 60 \pm 8 $ MeV, the scalar nucleon form 
factor at the Cheng-Dashen point $t=2M_\pi^2$, and the difference 
$\Delta_\sigma =\sigma_{\pi N}(2M_{\pi}^2) -\sigma_{\pi N}(0)=15.2 \pm 0.4$ 
MeV~\cite{Gasser:1990ce} 

as induced by explicit chiral symmetry breaking. 
The value of the isovector kaon-nucleon sigma-term $\sigma_{KN}^{I=1}$ 
was estimated in Ref.~\cite{Gasser:2000wv} using the baryon mass formulas: 
\begin{eqnarray}
\sigma_{KN}^{I=1} \sim \frac{m_s + \hat{m}}{m_s - \hat{m}} \, 
\frac{m_\Xi^2 - m_\Sigma^2}{8 m_P} = 48 \,\, {\rm MeV} 
 \sim 50 \,\,  {\rm MeV}  \, .
\end{eqnarray} 
Substituting the above values of the hadronic parameters to 
Eq.~(\ref{sigma}) we obtain for the scalar nucleon form factors:
\begin{eqnarray}\label{sffn}
G_{S}^{u} = 3.74\,, \ G_{S}^{d} = 2.69\,, \ G_{S}^{s} = 0.64\,.
\end{eqnarray}  
We have to point out that these values contain appreciable theoretical and 
experimental uncertainties, as seen from their derivation (for the possible 
error bars see, for instance, Ref.~\cite{Corsetti:2000yq}). In our analysis 
of the DNM contribution to \mbox{\m conversion} we take the numbers 
from Eq.~(\ref{alpha}) as central values of the scalar nucleon form factors. 
Note, in Ref.~\cite{Kosmas:2001mv} the different set of the values for 
the nucleon scalar form factors was derived on the basis of the QCD sum rules 
input parameters, which overestimates the pion-nucleon 
sigma-term and the strangeness of the nucleon. 
 
\section{Scalar Meson Contribution} 

In the following we turn to the scalar meson-exchange mechanism 
of \m conversion. Although the status of scalar mesons is still 
unclear~\cite{Eidelman:2004wy} we think it is reasonable to study their 
effect in the \m conversion since their contribution is associated with 
the experimentally most interesting coherent mode of this exotic process. 

The lightest unflavored scalar mesons are the isoscalar $f_0(600)$ and the 
isotriplet $a_0(980)$ states.
The former in the context of the nonlinear realization of chiral 
symmetry can be treated as a resonance in the $\pi\pi$ system (see detailed 
discussion, e.g. in Refs.~\cite{Colangelo:2001df,Oset:2000gn}). 
For simplicity we neglect a possible small strange content of the isoscalar 
meson and treat this state as $\bar u u + \bar d d$. 

We derive the LFV lepton-meson effective Lagrangian 
in terms of the interpolating $f_0$ and $a_0^0$ fields. Retaining all 
the interactions consistent with Lorentz invariance, we obtain the general 
form of this Lagrangian: 
\begin{eqnarray}\label{eff-LS}
{\cal L}_{eff}^{lS}\ &=& \  \frac{\Lambda_H^2}{\Lambda_{LFV}^2}  
\biggl[ ( \xi_S^{f_0} j^S \, + \, \xi_P^{f_0}j^P ) \, f_0 
      + ( \xi_S^{a_0}  j^S   \, + \, \xi_P^{a_0} j^P ) \, a_0^0 \biggr] 
\end{eqnarray}  
with the unknown dimensionless coefficients $\xi$ to be determined 
from the had\-ro\-ni\-za\-ti\-on prescription. 
We assume that the Lagrangian to be generated by the quark-lepton 
Lagrangian~(\ref{eff-q}), and, therefore, all its terms have the same 
suppression $\Lambda_{LFV}^{-2}$ with respect to the large LFV scale 
$\Lambda_{LFV}$. Another scale in the problem is the hadronic scale 
$\Lambda_H \sim 1$ GeV which adjusts the physical dimensions of the terms 
in  Eq.~(\ref{eff-LS}). In the Lagrangian in Eq.~(\ref{eff-LS}) we neglect 
derivative terms since their contribution to \m conversion is suppressed 
by a factor $(m_{\mu}/\Lambda_H)^2\sim 10^{-2}$. 

In order to relate the parameters $\xi$ of the Lagrangian~(\ref{eff-LS}) 
to the ``fundamental" parameters $\eta$ of the quark-lepton 
Lagrangian~(\ref{eff-q}) we use an approximate method based on the standard 
on-mass-shell matching condition~\cite{Faessler:1996ph}
\begin{equation}\label{match} 
\langle \mu^+ \, e^-|{\cal L}_{eff}^{lq}|S\rangle \approx 
\langle \mu^+ \, e^-|{\cal L}_{eff}^{lS}|S \rangle ,  
\end{equation} 
where $|S= f_0, a_0 \rangle$ are the on mass-shell scalar meson states.  
We solve equation~(\ref{match}) using the quark current matrix elements
\begin{eqnarray}\label{mat-el2} 
&&\langle 0|\bar u \, u|f_0(p)\rangle \, = \, 
  \langle 0|\bar d \, d|f_0(p)\rangle 
\, = \, m_{f_0}^2 \, f_{f_0}\,, 
\\ 
&&     \langle 0|\bar u \, u|a_0^0(p)\rangle = 
\,- \, \langle 0|\bar d \, d|a_0^0(p)\rangle 
\, = \, m_{a_0}^2 \, f_{a_0^0} \,. 
\nonumber
\end{eqnarray} 
Here $p$, $m_S$ and $f_{S}$ are the 
scalar-meson four-momentum, mass and decay constant, respectively. 
The quark operators in Eq.~(\ref{mat-el2}) are taken at $x=0$. 
In the numerical calculations we use the following values of scalar 
meson masses~\cite{Eidelman:2004wy}: 
\begin{eqnarray}\label{constants}  
m_{f_0} = 500 \,\,  \mbox{MeV}\,,  \hspace*{1cm} 
m_{a_0} = 984.7 \,\, \mbox{MeV}  \,. 
\end{eqnarray} 
The coupling constants $f_{S}$ in Eqs.~(\ref{mat-el2}) can 
be estimated using the linear $\sigma$-model in the case of 
the $f_0$ meson~\cite{Delbourgo:1993dk} and by QCD sum rules in the case 
of the $a_0$~\cite{Maltman:1999jn}. 
In the linear $\sigma$-model 
one has the following relationship ~\cite{Delbourgo:1993dk}:
\begin{eqnarray}\label{mat-el3}  
\langle 0|\bar u \, u|f_0(p)\rangle = m_{f_0}^2 \, \frac{\sqrt{N_c}}{2\pi}, 
\end{eqnarray} 
where $N_c=3$ is the number of quark colors.
Comparing Eqs.~(\ref{mat-el2}) and (\ref{mat-el3}) we get  
\begin{eqnarray}
f_{f_0} = \frac{\sqrt{N_c}}{2\pi} = 0.28 \,. 
\end{eqnarray} 
The coupling constant $f_{a_0^0}$ of the neutral $a_0^0$ meson is 
related to the coupling constant $f_{a_0^\pm}$ of the charged 
$a_0^\pm$ state due to isospin invariance: 
\begin{eqnarray}\label{a0-apm}
f_{a_0^\pm} \, =  \, f_{a_0^0} \, \sqrt{2}\,,
\end{eqnarray}  
with the definition $\langle 0|\bar d \, u|a_0^-(p)\rangle = 
m_{a_0}^2  f_{a_0^\pm}\,$.
The value of $f_{a_0^\pm}$ was estimated in Ref.~\cite{Maltman:1999jn}: 
\begin{eqnarray}\label{apm}
f_{a_0^\pm} = \frac{0.0447 \ {\rm GeV}^3}{m_{a_0}^2 \ (m_s - \hat{m})} \,. 
\end{eqnarray}
Combining Eqs.~(\ref{apm}) and~(\ref{a0-apm}) we have
\begin{eqnarray}
f_{a_0^0} \, = \, 0.19 \,. 
\end{eqnarray}
Solving Eq.~(\ref{match}) with the help of Eqs.~(\ref{mat-el2}), 
we obtain the expressions for the coefficients $\xi$ of the 
lepton-meson Lagrangian~(\ref{eff-LS}) in terms of the generic LFV  
parameters $\eta$ of the initial~(\ref{eff-q}) lepton-quark effective 
Lagrangian:  
\begin{eqnarray}
\xi_b^{a_0} \, = \, \left(\frac{m_{a_0}}{\Lambda_H}\right)^2 
f_{a_0^0} \,\eta_{b S}^{(3)}\,, \hspace*{1cm}  
\xi_b^{f_0} \, = \, \left(\frac{m_{f_0}}{\Lambda_H}\right)^2  
\, f_{f_0} \, \eta_{b S}^{(0)}\,, 
\end{eqnarray} 
where $b = S, P$ and $\eta^{(0,3)}=\eta^{u}\pm\eta^{d}$.

For our analysis we also need the effective Lagrangian describing the 
interactions of the scalar mesons with nucleons.
We take it in the following form: 
\begin{eqnarray}\label{MN} 
{\cal L}_{SN} \, = \, 
\bar{N} [ g_{_{a_0 NN}} \, \vec{a_0} \, 
\vec{\tau} \,  + \, g_{_{f_0 NN}} \, f_0] N\,. 
\end{eqnarray} 
For the meson-nucleon couplings $g_{SNN}$ we adopt the central values 
\begin{eqnarray}\label{SN-couplings}
g_{_{a_0 NN}} \simeq g_{_{f_0 NN}} \simeq 5 \, 
\end{eqnarray} 
used in phenomenological description of nucleon-nucleon interactions and 
recently also calculated in the chiral unitary approach~\cite{Oset:2000gn}. 
Since our analysis does not pretend to high accuracy we do not supply 
the error bars for the meson-nucleon couplings $g_{SNN}$. 

Now, having specified the interactions of the scalar mesons $f_0, a_0$ with 
leptons and with nucleons we 
can derive the scalar meson-exchange contributions to the \m conversion. 
This contribution can be expressed in the form of the nucleon-lepton effective 
Lagrangian (\ref{eff-N}) which arises in second order in the Lagrangian 
${\cal L}_{eff}^{lS} + {\cal L}_{SN}$ and corresponds to the diagram 
in Fig.1b. We estimate this contribution only for the coherent 
$\mu^- - e^-$ conversion process. In this case we disregard all the 
derivative terms of nucleon and lepton fields. Neglecting the kinetic 
energy of the final nucleus, the muon binding energy and the electron 
mass, the square of the momentum transfer $q^2$ to the nucleus has 
a constant value $q^2 \approx - m_{\mu}^2$. In this approximation the 
meson propagators convert to $\delta$-functions leading to effective 
lepton-nucleon contact type operators. Comparing them with the 
corresponding terms in the Lagrangian (\ref{eff-N}), we obtain the scalar 
meson exchange contribution to the coupling constants of this Lagrangian: 
\begin{eqnarray} \label{alpha-S-ex}
\alpha_{bS[MEM]}^{(3)} &=& \beta_{a_0}\eta_{bS}^{(3)}\, 
\hspace*{1cm}  
\alpha_{bS[MEM]}^{(0)} = \beta_{f_0}\eta_{bS}^{(0)}
\end{eqnarray}  
with $b=S,P$ and the coefficients 
\begin{eqnarray} \label{beta}
\beta{_{a_0}} = \frac{g_{_{a_0 NN}} \, f_{a_0^0} \, 
m_{a_0}^2}{m_{a_0}^2 
+ m_{\mu}^2}\,, \hspace*{1cm} 
\beta{_{f_0}} = \frac{g_{_{f_0 NN}} \, f_{f_0} \, m_{f_0}^2 }
{m_{f_0}^2 + m_{\mu}^2}\,.
\end{eqnarray}
Substituting the values of the meson coupling constants and masses  
we obtain for these coefficients 
\begin{eqnarray} \label{beta-num} 
\beta{_{a_0}} = 0.93 \,, \hspace*{1cm} \beta{_{f_0}} = 1.32 \,,
\end{eqnarray} 
These values should be considered as rough estimates in view of the 
uncertainties in the scalar meson masses and couplings. 

\section{Constraints on LFV parameters from \m conversion}

 From the Lagrangian~(\ref{eff-N}), following the standard procedure, 
one can derive the formula for the branching ratio of the coherent 
$\mu^- - e^-$ conversion. To leading order in the nonrelativistic 
reduction the branching ratio takes the form~\cite{Kosmas:1993ch}
\begin{equation} 
R_{\mu e^-}^{coh} \ = \  
\frac{{\cal Q}} {2 \pi \Lambda_{LFV}^4} \  \   
\frac{p_e E_e \ ({\cal M}_p + {\cal M}_n)^2 } 
{ \Gamma_{\mu c} } 
\, , 
\label{Rme}
\end{equation} 
where $p_e, E_e$ are 3-momentum and energy of the outgoing electron,  
${{\cal M}}_{p,n}$ are the nuclear $\mu^- - e^-$  transition matrix elements 
and $\Gamma_{\mu c}$ is the total rate of the ordinary muon capture. 
The factor ${\cal Q}$ takes the form 
\begin{eqnarray}
\hspace*{-1cm}
{\cal Q} &=& |\alpha_{VV}^{(0)}+\alpha_{VV}^{(3)}\ \phi|^2 +
|\alpha_{AV}^{(0)}+\alpha_{AV}^{(3)} \phi|^2 + 
|\alpha_{SS}^{(0)}+\alpha_{SS}^{(3)} \phi|^2 + 
|\alpha_{PS}^{(0)} + \alpha_{PS}^{(3)} \phi|^2 
\nonumber  \\ 
\hspace*{-1cm} 
&+& 2{\rm Re}\{(\alpha_{VV}^{(0)}+\alpha_{VV}^{(3)} 
\phi)(\alpha_{SS}^{(0)}+ \alpha_{SS}^{(3)} \phi)^\ast
+  (\alpha_{AV}^{(0)}+\alpha_{AV}^{(3)}\ \phi)(\alpha_{PS}^{(0)} + 
\alpha_{PS}^{(3)}\ \phi)^\ast\}\,
\label{Rme.1} 
\end{eqnarray} 
in terms of the parameters of the lepton-nucleon effective 
Lagrangian (\ref{eff-N}) and the nuclear structure factor 
\begin{eqnarray}\label{phi}
\phi = ({\cal M}_p - {\cal M}_n)/({\cal M}_p + {\cal M}_n) \, ,
\end{eqnarray}
that is typically small for the experimentally interesting nuclei.

The nuclear matrix elements ${\cal M}_{p,n}$ have been calculated in 
Refs.~\cite{Kosmas:2001mv,Faessler:pn} for the nuclear targets 
$^{27}$Al, $^{48}$Ti and $^{197}$Au. We show their values in Table~II 
together with the experimental values of the total rates $\Gamma_{\mu c}$ of 
the ordinary muon capture~\cite{Suzuki:1987jf} and the 3-momentum $p_e$ of 
the outgoing electron. Using the quantities from Table II we find for the 
dimensionless scalar lepton-nucleon couplings of the Lagrangian~(\ref{eff-N}) 
the following limits
\begin{eqnarray}\label{alpha-lim}
\alpha_{bS}^{(k)} \left(\frac{1 \mbox{GeV}}{\Lambda_{LFV}}\right)^2 
\leq 1.2 \times 10^{-12} \left[B^{(k)}(A)\right]^{-2}, \ \ \ 
\end{eqnarray}
with $k=0,3$.
Here the scaling factors $B^{(k)}(A)$ depend on the target nucleus $A$ used 
in an experiment setting the upper limit 
$R_{\mu e}^{A}(Exp)$  on the branching ratio of \m conversion: 
$R_{\mu e}^{A}\leq R_{\mu e}^{A}(Exp)$. The numerical values of these factors 
for the experiments discussed in the introduction have been calculated with 
the help of Table~II and are  given in Table~I.

From the limits in Eq.~(\ref{alpha-lim}) one can deduce individual limits on 
the terms contributing to the coefficients $\alpha_{bS}^{(0,3)}$, assuming 
the absence of significant cancellations (unnatural fine-tuning) between 
the different terms. In this way, using Eqs.~(\ref{alpha}), 
(\ref{alpha-S-ex}), we derive constraints on the 4-fermion quark-lepton 
LFV couplings of the Lagrangian~(\ref{eff-q}) for the two considered 
mechanisms of hadronizations: the direct nucleon mechanism (DNM) and the 
meson-exchange mechanism (MEM). The corresponding limits are shown in Table~3. 
Following the common practice, 
we presented these limits in terms of the individual mass scales, 
$\Lambda_{ij}$, of the scalar quark-lepton contact operators in 
Eq.~(\ref{eff-q}). In the conventional definition~\cite{Eichten:1983hw} 
these scales determine the couplings 
$z_{ij}=g^2/\Lambda^2_{ij}$ of the 4-fermion contact terms of the form 
$z_{ij} (\bar{l_{i}}l_j)(\bar{q}q)$ with a fixed $g^2=4\pi$. 
Historically~\cite{Eichten:1983hw} this definition of $\Lambda_{ij}$ 
originates from substructure models and corresponds to 
the compositeness scale in the strong coupling regime $g^2/4\pi=1$.
Thus, the individual mass scales $\Lambda_{ij}$, introduced in this way, 
are related to our notations according to the formula:
\begin{eqnarray}\label{Lambda-LFV}
\frac{\eta_{bS}^{(k)}}{\Lambda^2_{LFV}} = 
\frac{4\pi}{\left(\Lambda^{(k,b)}_{\mu e}\right)^2}
\end{eqnarray}
with $k=0, 3, s$ and $b=P,S$.  

Let us compare our limits for these mass scales with the corresponding 
limits existing in the literature.  
The limits  on $\Lambda_{\mu e}$ can be derived from the experimental bounds 
on the rates of the $\pi^+\rightarrow \mu^+ \nu_e$, $\pi^0\rightarrow 
\mu^\pm e^\mp$ decays~\cite{Kim:1997rr}. Despite the scalar quark current 
does not directly contribute to these processes the limits come from 
the gauge invariance with respect to the SM group which relates 
the couplings of the scalar and pseudoscalar lepton-quark contact operators. 
Typical limits from these processes are of 
$\Lambda_{\mu e} \geq~\mbox{few TeV}$.
In the future experiments at the LHC it is also planned to set limits on 
the mass scales of various contact quark-lepton interactions 
from the measurement of Drell-Yan cross sections in the high dilepton mass 
region~\cite{Krasnikov:2003ef,Gupta:1999iy}. 
In this case typical expected limits for the scales of the lepton flavor 
diagonal contact terms are $\Lambda_{l l} \geq 35~\mbox{TeV}$. 
We are not aware of the corresponding analysis for the limits on the 
LFV scales, like $\Lambda_{\mu e}$, expected from the experiments planned 
at the LHC. However, one may expect these limits to be significantly 
stronger (typically by an order of magnitude) than the above cited limits 
for the lepton flavor diagonal mass scales $\Lambda_{l l}$. 
This is motivated by the fact that, usually a flavor diagonal process 
have much more SM background than the LFV processes. 
A comparison of the above limits with the limits in Table~III, extracted 
from \mbox{\m conversion}, shows that the latter are more stringent.  

The following comment is in order. As follows from Eq.~(\ref{Rme.1}) 
and Table~II, the contribution of the isovector couplings $\alpha^{(3)}$ 
to the \m conversion rate, Eqs.~(\ref{Rme}) and~(\ref{Rme.1}), is suppressed 
by a small nuclear factor $\phi$. On the other hand, the information on the 
isovector couplings may be important for the phenomenology of the physics 
beyond the SM allowing one to distinguish the contribution of d and u quarks. 
In this respect, the contribution of the scalar mesons is important,  
since it is comparable on magnitude with the DNM mechanism. Taking this 
contribution into account we extracted reasonable constraints on mass scales 
of the isovector quark-lepton contact terms $\Lambda^{(3)}_{\mu e}$ presented 
in Table~3. 
The strange quark contributions are not affected by the meson exchange 
because we neglected the possible strange quark component of the scalar mesons.

\section{Summary} 

We analyzed the nuclear $\mu^--e^-$ conversion in the general framework 
the effective Lagrangian approach without referring to any specific 
realization of the physics beyond the standard model responsible 
for lepton flavor violation. 
The two mechanisms of the hadronization of the underlying effective 
quark-lepton LFV Lagrangian have been studied: the direct nucleon (DNM) and 
the scalar meson-exchange (MEM) mechanisms. 
We showed that the scalar meson-exchange contribution is comparable on 
magnitude with the DNM mechanism and, therefore, can modify the limits 
on the LFV lepton-quark couplings derived on the basis of the conventional 
direct nucleon mechanism. 
These results lead us to the conclusion that the meson exchange mechanism 
may have an appreciable impact on the phenomenology of the LFV physics 
beyond the standard model and, therefore, should be taken in to account in 
the analysis of the LFV effects in hadronic and nuclear semileptonic processes.

From the experimental upper bounds on the \m conversion rate we extracted 
the lower limits on the mass scales of the LFV lepton-quark contact terms 
involved in this process and showed that they are more stringent than 
the similar limits existing in the literature.
 
\vspace*{1cm}

{\bf Acknowledgments}

\noindent 
This work was supported by the FONDECYT projects 1030244, 1030355,  
by the DFG under contracts FA67/25-3, GRK683.  
This research is also part of the EU Integrated Infrastructure
Initiative Had\-ron\-phy\-si\-cs project under contract number
RII3-CT-2004-506078 and President grant of Russia "Scientific Schools" 
No.1743.2003.

\newpage

\begin{table}
{\bf Table I.} The experimental upper limits on the 
\m conversion branching ratio $R_{\mu e}^{A}$ and the values of the 
scaling factors $B^{(0,3)}(A)$ from Eq.~(\ref{alpha-lim})
 for the running and forthcoming experiments. 

\vspace*{.4cm} 

\begin{center} 
\begin{tabular}{|l|l|c|c|} 
Target, Experiment &$R_{\mu e}^{A}$ &$B^{(0)}(A)$ &$B^{(3)}(A)$\\ 
\hline 
&&&    \\ 
$^{48}$Ti, SINDRUM II\cite{Honecker:1996zf}& $6.1\times 10^{-13}$(90\% C.L.) 
& 1 &0.3\\  
&&&    \\ 
$^{27}$Al, MECO\cite{MECO} &$\sim 2\times 10^{-17}$ $^{\dagger}$ 
&11.5 & 1.7\\ 
&&&    \\ 
$^{197}$Au, SINDRUM II\cite{Honecker:1996zf}
&$\sim 6\times 10^{-13}$ $^{\dagger}$ & 1.27 & 0.46\\ 
&&&    \\ 
$^{48}$Ti, PRIME\cite{PRIME}&$\sim 10^{-18}$ $^{\dagger}$
& 28 & 8.8\\ 
&&&    \\ 
\end{tabular}
\end{center} 
$^{\dagger}$ expected upper limits.
\end{table}

\begin{table}
{\bf Table II.} Transition nuclear matrix elements ${\cal M}_{p,n}$ 
from Eqs. (\ref{Rme}), (\ref{phi}) and 
other useful quantities
(see the text).

\vspace*{.4cm} 

\begin{center}
\begin{tabular}{|r|c|c|c|c|}
   &  &  &  &    \\
Nucleus & $p_e \, (fm^{-1})$ &
$\Gamma_{\mu c} \, ( \times 10^{6} \, s^{-1})$ &
${\cal M}_p(fm^{-3/2})$ & ${\cal M}_n(fm^{-3/2})$  \\
\hline
   &  &  &  &    \\
$^{27}$Al  & 0.531 &   0.71 & 0.047 & 0.045   \\
   &  &  &  &   \\
$^{48}$Ti  & 0.529 &  2.60 & 0.104 & 0.127   \\
    &  &  &  &  \\
$^{197}$Au & 0.485& 13.07 & 0.395 & 0.516   \\
\end{tabular}
\end{center}
\end{table}

\begin{table}
{\bf Table III.} Lower limits on the individual mass scales, 
$\Lambda_{\mu e}$, of the scalar quark-lepton contact operators 
in Eq.~(\ref{eff-q}) inferred from the experimental upper bounds on 
the branching ratio of \m conversion for the direct nucleon mechanism (DNM) 
and the meson exchange mechanism (MEM). 
The superscript notation is $b = P,S$. The values of the scaling factors 
$B^{(0,3)}(A)$ for the running and some planned \m conversion experiments 
are given in Table 2.

\vspace*{.4cm} 

\begin{center} 
\begin{tabular}{|c|c|c|} 
LFV Mass Scale & DNM \hspace*{1.5cm}& MEM \hspace*{1.5cm}\\ 
\hline 
&&    \\ 
$\Lambda^{(0,b)}_{\mu e}$
& $5.8\times 10^{3} B^{(0)}(A)$ TeV \hspace*{1.5cm}
& $3.7\times 10^{3} B^{(0)}(A)$ TeV \hspace*{1.5cm}\\
&&    \\ 
$\Lambda^{(3,b)}_{\mu e}$
& $2.3\times 10^{3} B^{(3)}(A)$ TeV \hspace*{1.5cm}
& $3.1\times 10^{3} B^{(3)}(A)$ TeV \hspace*{1.5cm}\\
&&    \\ 
$\Lambda^{(s,b)}_{\mu e}$
& $2.6\times 10^{3} B^{(0)}(A)$ TeV\hspace*{1.5cm}
& no limits \hspace*{1.5cm}\\
&&    \\ 
\end{tabular}
\end{center} 
\end{table}

\newpage 
 
\begin{figure}
 
\noindent {\bf Fig.1:}
Diagrams contributing to the nuclear $\mu^--e^-$ conversion in 
the scalar channel: 
direct nucleon mechanism (a) and
meson-exchange mechanism (b).
\end{figure}
 
\begin{figure}[t]
                                                                               
\vspace*{2cm}
 
\centering{\epsfxsize=15 cm\epsffile{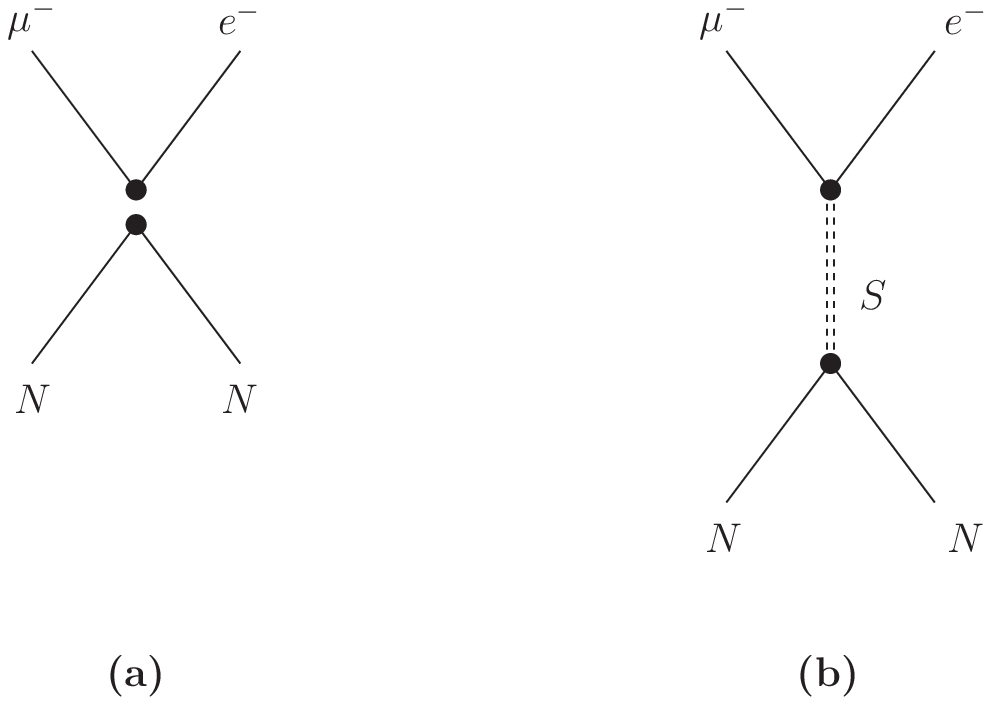}}
 
\vspace*{.5cm}
 
\centerline{\bf Fig.1}
\end{figure}
 

\begin{thebibliography}{99}  
\bibitem{Kosmas:1993ch}
  T.~S.~Kosmas, G.~K.~Leontaris and J.~D.~Vergados,
  Prog.\ Part.\ Nucl.\ Phys.\  {\bf 33}, 397 (1994)
  [arXiv:hep-ph/9312217].
\bibitem{mu-e theory-exp} 
W.J. Marciano, Lepton flavor violation, 
summary and perspectives, Summary talk in the conference on 
"New initiatives in lepton flavor violation and neutrino oscillations 
with very intense muon and neutrino beams", Honolulu-Hawaii, USA, 
Oct. 2-6, 20002, River Edge, NJ: World Scientific, 2000 
(see also $http://meco.ps.uci.edu/lepton\_workshop.$);  
Y.~Kuno and Y.~Okada,
Rev.\ Mod.\ Phys.\  {\bf 73}, 151 (2001)
[arXiv:hep-ph/9909265].
\bibitem{Honecker:1996zf} 
W.~Honecker {\it et al.} [SINDRUM II Collaboration] ,
Phys.\ Rev.\ Lett.\  {\bf 76}, 200 (1996);  
P. Wintz, in Status of muon to electron conversion at PSI,  
Invited talk  at~\cite{mu-e theory-exp}; 
A.~van~der Schaaf, Private communication.  
\bibitem{Molzon}W.~Molzon, The improved tests of muon and electron 
flavor symmetry in muon processes, Spring. Trac. Mod. Phys. 
{\bf 163}, 105 (2000). 
\bibitem{MECO}
J.~Sculli, The MECO experiment, 
Invited talk at~\cite{mu-e theory-exp}. 
\bibitem{PRIME}
Y.~Kuno, The PRISM: Beam-Experiments, 
Invited talk  at~\cite{mu-e theory-exp}. 
\bibitem{Faessler:2004ea} 
A.~Faessler, T.~Gutsche, S.~Kovalenko, V.~E.~Lyubovitskij, 
I.~Schmidt and F.~Simkovic, 
Phys.\ Rev.\ D {\bf 70} (2004) 055008 
[arXiv:hep-ph/0405164].  
\bibitem{Kosmas:2001mv}
T.~S.~Kosmas, S.~Kovalenko and I.~Schmidt,
Phys.\ Lett.\ B {\bf 511}, 203 (2001); 
Phys.\ Lett.\ B {\bf 519}, 78 (2001).
\bibitem{Faessler:pn}
A.~Faessler, T.~S.~Kosmas, S.~Kovalenko and J.~D.~Vergados,
Nucl.\ Phys.\ B {\bf 587}, 25 (2000). 
\bibitem{Faessler:2004jt}
A.~Faessler, T.~Gutsche, S.~Kovalenko, V.~E.~Lyubovitskij, 
I.~Schmidt and F.~Simkovic, 
Phys.\ Lett.\ B {\bf 590} (2004) 57 
[arXiv:hep-ph/0403033]. 
\bibitem{Inoue:2003bk}  
T.~Inoue, V.~E.~Lyubovitskij, T.~Gutsche and A.~Faessler, 
Phys.\ Rev.\ C {\bf 69} (2004) 035207 
[arXiv:hep-ph/0311275].
\bibitem{Gasser:1982ap}
J.~Gasser and H.~Leutwyler,
Phys.\ Rept.\  {\bf 87}, 77 (1982). 
\bibitem{Gasser:1990ce} 
J.~Gasser, H.~Leutwyler and M.~E.~Sainio,  
Phys.\ Lett.\ B {\bf 253} (1991) 252, 260. 
\bibitem{Gasser:2000wv}
J.~Gasser and M.~E.~Sainio,
arXiv:hep-ph/0002283; 
M.~E.~Sainio,
PiN Newslett.\  {\bf 16}, 138 (2002)
[arXiv:hep-ph/0110413]. 
\bibitem{Corsetti:2000yq} 
A.~Corsetti and P.~Nath, 
Phys.\ Rev.\ D {\bf 64}, 125010 (2001) 
[arXiv:hep-ph/0003186].
\bibitem{Eidelman:2004wy}
S.~Eidelman {\it et al.} [Particle Data Group Collaboration], 
Phys.\ Lett.\ B {\bf 592}, 1 (2004). 
\bibitem{Colangelo:2001df} 
G.~Colangelo, J.~Gasser and H.~Leutwyler, 
Nucl.\ Phys.\ B {\bf 603} (2001) 125 
[arXiv:hep-ph/0103088];  
J.~A.~Oller, E.~Oset and J.~R.~Pelaez, 
Phys.\ Rev.\ Lett.\  {\bf 80}, 3452 (1998) 
[arXiv:hep-ph/9803242]. 
\bibitem{Oset:2000gn}
E.~Oset, H.~Toki, M.~Mizobe and T.~T.~Takahashi,
Prog.\ Theor.\ Phys.\  {\bf 103}, 351 (2000) 
[arXiv:nucl-th/0011008].
\bibitem{Faessler:1996ph}
A.~Faessler, S.~Kovalenko, F.~Simkovic and J.~Schwieger,
Phys.\ Rev.\ Lett.\  {\bf 78}, 183 (1997).
\bibitem{Delbourgo:1993dk} 
R.~Delbourgo and M.~D.~Scadron, 
Mod.\ Phys.\ Lett.\ A {\bf 10}, 251 (1995) 
[arXiv:hep-ph/9910242];  
R.~Delbourgo, M.~D.~Scadron and A.~A.~Rawlinson,
Mod.\ Phys.\ Lett.\ A {\bf 13}, 1893 (1998) 
[arXiv:hep-ph/9807505]. 
\bibitem{Maltman:1999jn} 
K.~Maltman, 
Phys.\ Lett.\ B {\bf 462}, 14 (1999)
[arXiv:hep-ph/9906267].
\bibitem{Suzuki:1987jf} 
T.~Suzuki, D.~F.~Measday and J.~P.~Roalsvig,
Phys.\ Rev.\ C {\bf 35}, 2212 (1987). 
\bibitem{Eichten:1983hw}
  E.~Eichten, K.~D.~Lane and M.~E.~Peskin,
  Phys.\ Rev.\ Lett.\  {\bf 50}, 811 (1983); 
V.~D.~Barger, K.~Cheung, K.~Hagiwara and D.~Zeppenfeld,
Phys.\ Lett.\ B {\bf 404}, 147 (1997)  
[arXiv:hep-ph/9703311];  
Phys.\ Rev.\ D {\bf 57}, 391 (1998)
[arXiv:hep-ph/9707412].
\bibitem{Kim:1997rr}
  J.~E.~Kim, P.~Ko and D.~G.~Lee,
  Phys.\ Rev.\ D {\bf 56}, 100 (1997)
  [arXiv:hep-ph/9701381].
\bibitem{Krasnikov:2003ef}
  N.~V.~Krasnikov and V.~A.~Matveev,
  Phys.\ Usp.\  {\bf 47}, 643 (2004)
  [Usp.\ Fiz.\ Nauk {\bf 174}, 697 (2004)]
  [arXiv:hep-ph/0309200].
\bibitem{Gupta:1999iy}
  A.~K.~Gupta, N.~K.~Mondal and S.~Raychaudhuri,
  arXiv:hep-ph/9904234.  
\end{thebibliography}
\end{document}